\documentclass{article}
\usepackage{spconf,amsmath,amssymb,graphicx,url}
\usepackage[breaklinks,colorlinks]{hyperref}

\title{VoiceLDM: Text-to-Speech with Environmental Context}
\name{Yeonghyeon Lee, Inmo Yeon, Juhan Nam, Joon Son Chung
\thanks{This work was supported by the National Research Foundation of Korea (NRF) grant funded by the Korea government (MSIT) (No. RS-2023-00222383).}
}
\address{
Korea Advanced Institute of Science and Technology, South Korea
}
\begin{document}
\ninept
\maketitle
\begin{abstract}
This paper presents VoiceLDM, a model designed to produce audio that accurately follows two distinct natural language text prompts: the description prompt and the content prompt. The former provides information about the overall environmental context of the audio, while the latter conveys the linguistic content. To achieve this, we adopt a text-to-audio (TTA) model based on latent diffusion models and extend its functionality to incorporate an additional content prompt as a conditional input. By utilizing pretrained contrastive language-audio pretraining (CLAP) and Whisper, VoiceLDM is trained on large amounts of real-world audio without manual annotations or transcriptions. Additionally, we employ dual classifier-free guidance to further enhance the controllability of VoiceLDM. Experimental results demonstrate that VoiceLDM is capable of generating plausible audio that aligns well with both input conditions, even surpassing the speech intelligibility of the ground truth audio on the AudioCaps test set. Furthermore, we explore the text-to-speech (TTS) and zero-shot text-to-audio capabilities of VoiceLDM and show that it achieves competitive results. Demos and code are available at \url{https://voiceldm.github.io}.

% **from joon**: it would be nice to have a small 'teaser figure' on the first page - showing an example like two inputs going into the model, and an output (sound symbol)

% The abstract should contain about 100 to 150
% words, and should be identical to the abstract text submitted electronically
% along with the paper cover sheet.
\end{abstract}
\section{Introduction}
\label{sec:intro}

Recent advances in text-to-audio (TTA) generation have shown impressive performance in terms of fidelity and diversity~\cite{liu2023audioldm, kreuk2023audiogen, huang2023make, huang2023make2, yang2023diffsound, borsos2023audiolm, ghosal2023text}. These models demonstrate the ability to synthesize audio that accurately reflects the semantic context provided by a natural language prompt.
Nevertheless, one limitation of these models is that when prompted to produce speech (e.g. \textit{``a man is speaking in a cathedral"}), instead of generating audio with coherent linguistic output, they often generate incoherent babbling voices.

Motivated by this, we introduce VoiceLDM, a text-to-speech (TTS) model inspired by TTA models that also generate linguistically intelligible voices. VoiceLDM can be controlled with two types of natural language prompts, a content prompt specifying the linguistic content of the spoken utterance, and a description prompt that characterizes the environmental context of the audio. Our work can be seen as standing at the intersection of text-to-speech and text-to-audio. To the best of our knowledge, it is the first work that simultaneously achieves the speech intelligibility present in TTS models while also having the diverse audio generation capability found in TTA models. As a result, our model is capable of generating a wide range of sounds, such as speech with sound effects, singing voices, whispering, and more.

There have been recent or concurrent works in TTS which also possesses the capability to utilize a second text prompt to control the style of the audio being generated~\cite{liu2023audioldm2, guo2023prompttts, yang2023instructtts, liu2023promptstyle, kim21n_interspeech}. However, the controllable diversity is only limited to speech-related factors such as gender, emotion, and volume.

We build upon the work of AudioLDM \cite{liu2023audioldm}, a TTA system based on latent diffusion models. We extend the model by integrating an additional content prompt as a conditional input.
We train our model using real-world audio data by taking advantage of contrastive language-audio pretraining (CLAP) \cite{wu2023large} and Whisper \cite{radford2023robust}. We are thereby able to use large-scale audio datasets without human annotation, which are then used for model training to achieve better generation results.

Experimental results demonstrate that VoiceLDM generates audio that aligns well with both the content prompt and the description prompt. Furthermore, the audio generated by VoiceLDM often surpasses the linguistic intelligibility of the ground truth audio. We also show that the model is capable of functioning as a regular TTS or TTA model and demonstrate that it achieves competitive results on each task.

\begin{figure}[!t]
  % \vspace{+0.5cm}
  \includegraphics[width=0.48\textwidth]{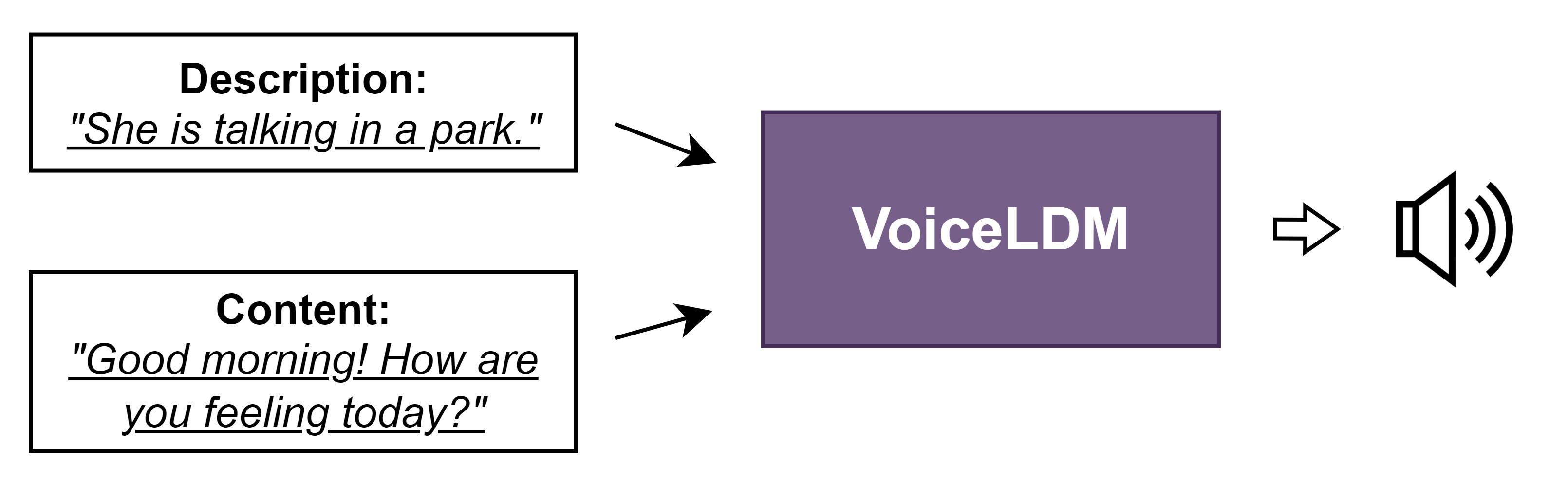}
  \caption{VoiceLDM produces audio that follows both the description prompt and the content prompt, bridging the gap between the domains of text-to-speech and text-to-audio.}
  % \vspace{-0.25cm}
  \label{fig:teasefig}
\end{figure}

\section{Method}
\label{sec:method}

\begin{figure*}[h]
  \vspace{-0.5cm}
  \centering
  \includegraphics[width=0.9\textwidth]{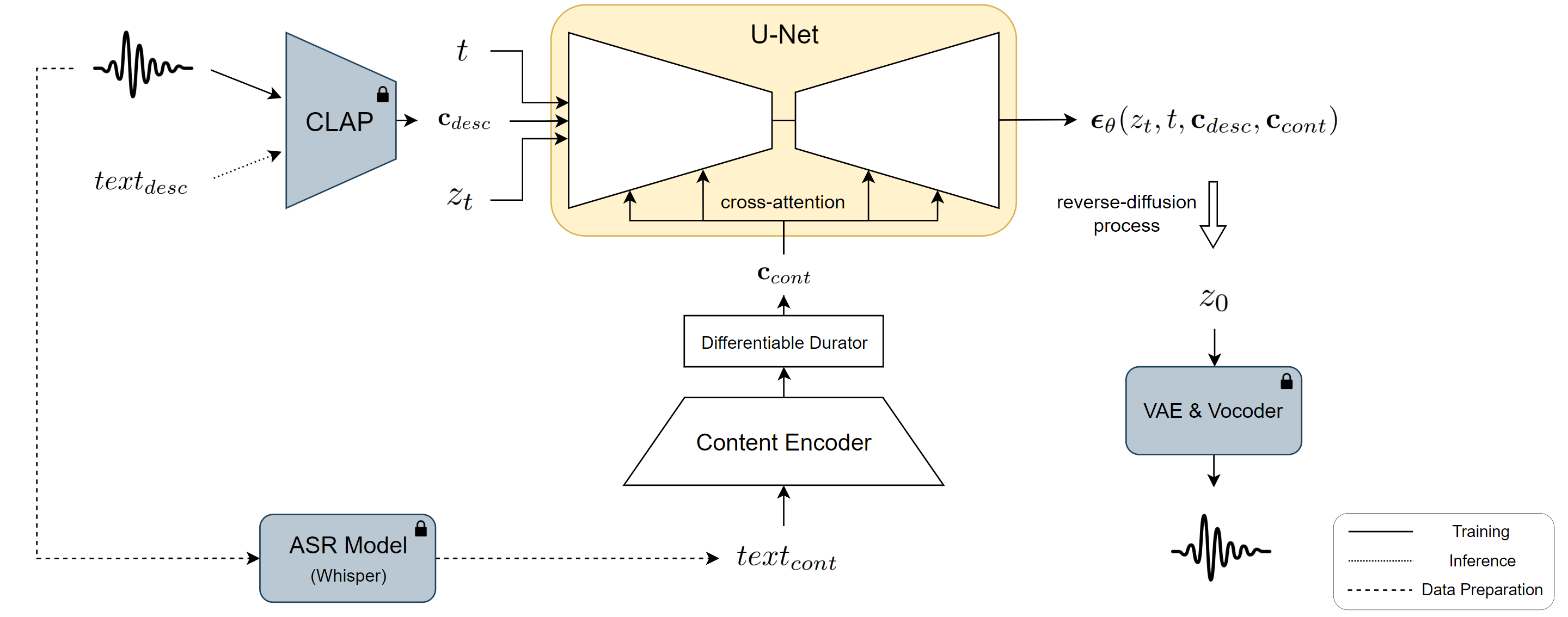}
  \caption{Overview of VoiceLDM. VoiceLDM is trained with large amounts of real-world audio data. $text_{cont}$ is generated during data preparation by processing the audio with Whisper, an automatic speech recognition (ASR) model. $text_{desc}$ is only used during inference. Modules with a lock icon indicates that it is frozen during training.}
  \vspace{-0.25cm}
  \label{fig:mainfig}
\end{figure*}

\subsection{Model Overview}
Figure \ref{fig:mainfig} illustrates the overall framework of VoiceLDM. Given two natural language text prompts $text_{cont}$ and $text_{desc}$, the role of VoiceLDM is to generate audio $\mathbf{X}$ that follows both conditions as input. The description prompt $text_{desc}$ is first converted into a 512-dimensional vector $\mathbf{c}_{desc} \in \mathbb{R}^{512}$ by the pre-trained CLAP \cite{wu2023large} model. A reference audio may also be used to attain $\mathbf{c}_{desc}$, since CLAP is designed to project both modalities into the same latent space.  The content prompt $text_{cont}$ is encoded into a hidden sequence $\mathbf{H}_{cont} \in \mathbb{R}^{L \times D}$ by the content encoder, where $L$ is the sequence length and $D$ is the dimension size. The differentiable durator then upsamples the hidden sequence into $\mathbf{c}_{cont} \in \mathbb{R}^{N \times D}$, where $L \leq N$. The differentiable durator is identical to the one used in \cite{tan2022naturalspeech}. The U-Net backbone \cite{ronneberger2015u} parameterized as $\theta$ takes in both conditions $\mathbf{c}_{desc}$ and $\mathbf{c}_{cont}$ and the timestep embedding to predict the diffusion score $\boldsymbol{\epsilon}_\theta$. 

% How is it conditioned to the U-Net?

Starting from a random noise sampled from an isotropic Gaussian distribution $z_{T} \sim \mathcal{N}(\textbf{0}, \textbf{I})$, the reverse diffusion process iteratively denoises $z_t$ for each time step $t$ with the predicted diffusion score $\boldsymbol{\epsilon}_{\theta}$ and predicts the initial audio prior $z_0$. $z_0$ can then be decoded back to the corresponding mel-spectrogram by the pre-trained variational autoencoder (VAE). Finally, the pre-trained HiFi-GAN vocoder \cite{kong2020hifi} converts the mel-spectrogram into desired audio $\mathbf{X}$. 

\subsection{Training}
The training procedure of VoiceLDM mostly follows the latent diffusion model training procedure as done in \cite{liu2023audioldm, rombach2022high, ho2020denoising}. However, the main difference is that the diffusion model utilizes two conditions. Starting from an audio $\mathbf{X}$, the pre-trained VAE compresses the audio into the latent representation $z_0$. A noisy representation of $z_0$ at a certain timestep $z_t$ is obtained by applying noise to $z_0$ through the forward diffusion process, following a predefined noise schedule.

Due to CLAP, manually annotated description prompt $text_{desc}$ is not necessary to obtain $\mathbf{c}_{desc}$ during training. Instead, CLAP is able to take in the original audio $\mathbf{x}$ to obtain the descriptive condition $\mathbf{c}_{desc}$. The content encoder and the differentiable durator encodes the speech transcription $text_{cont}$ into the content condition $\mathbf{c}_{cont}$. Finally, the model is trained to predict the added noise $\boldsymbol{\epsilon}$ with the following re-weighted training objective:

\begin{equation}
    \mathcal{L}_{\theta} = 
    % \mathop\mathbb{E}_{z_0, \epsilon, t}  
    \| \boldsymbol{\epsilon} - \boldsymbol{\epsilon}_\theta(z_t, t, \mathbf{c}_{desc}, \mathbf{c}_{cont}) \|_2^2
\end{equation}

Parameters of the U-Net backbone, the content encoder and the differentiable durator are all jointly trained. The pre-trained CLAP model, the pre-trained VAE and vocoder are kept frozen during training.

\subsection{Dual Classifier-Free Guidance}

An interesting property of VoiceLDM is that classifier-free guidance \cite{ho2021classifier} for the reverse diffusion process can be applied independently with respect to each condition $\mathbf{c}_{desc}$ and $\mathbf{c}_{cont}$. This allows one to trade-off mode coverage and sample fidelity for each individual conditions, allowing increased levels of controllability during generation \cite{cho2023towards}.
% Similar practice has been explored in 
When two conditions ($\mathbf{c}_{desc}$ and $\mathbf{c}_{cont}$) are viewed as one unified condition, it is possible to apply classifier-free guidance as follows:

% \begin{equation}
% \label{eq:cfg}
%     \boldsymbol{\Tilde{\epsilon}}(z_t, c_{desc}, c_{cont}) = \boldsymbol{\epsilon}(z_t, \O) + w(\boldsymbol{\epsilon}(z_t, c_{desc}, c_{cont}) - \boldsymbol{\epsilon}(z_t, \O))
% \end{equation}
\vspace{-0.25cm}
\begin{multline}
\label{eq:cfg}
    \boldsymbol{\Tilde{\epsilon}}_{\theta}(z_t, \mathbf{c}_{desc}, \mathbf{c}_{cont}) = \boldsymbol{\epsilon}_{\theta}(z_t, \mathbf{c}_{desc}, \mathbf{c}_{cont}) \\
    + w\Bigl(\boldsymbol{\epsilon}_{\theta}(z_t, \mathbf{c}_{desc}, \mathbf{c}_{cont}) - \boldsymbol{\epsilon}_{\theta}(z_t, \emptyset)\Bigr)
\end{multline}
where $w$ is the guidance strength and $\emptyset$ indicates the null condition. However, additional control can be achieved by applying dual classifier-free guidance. In this case, the diffusion score $\Tilde{\boldsymbol{\epsilon}}$ is formulated as follows:

% \begin{equation}
% \label{eq:cfg2}
%     \Tilde{\boldsymbol{\epsilon}}(z_t, c_{desc}, c_{cont}) = \boldsymbol{\epsilon}(z_t, c_{desc}, c_{cont})
%     \\+ w_{desc}(\boldsymbol{\epsilon}(z_t, c_{desc}, \O_{cont}) - \boldsymbol{\epsilon}(z_t, \O_{desc}, \O_{cont}))
%     + w_{cont}(\boldsymbol{\epsilon}(z_t, \O_{desc}, c_{cont}) - \boldsymbol{\epsilon}(z_t, \O_{desc}, \O_{cont}))
% \end{equation}
\vspace{-0.25cm}
\begin{multline}
\label{eq:cfg2}
    \Tilde{\boldsymbol{\epsilon}}_{\theta}(z_t, \mathbf{c}_{desc}, \mathbf{c}_{cont}) = \boldsymbol{\epsilon}_{\theta}(z_t, \mathbf{c}_{desc}, \mathbf{c}_{cont}) \\
    + w_{desc}\Bigl(\boldsymbol{\epsilon}_{\theta}(z_t, \mathbf{c}_{desc}, \emptyset_{cont}) - \boldsymbol{\epsilon}_{\theta}(z_t, \emptyset_{desc}, \emptyset_{cont})\Bigr) \\
    + w_{cont}\Bigl(\boldsymbol{\epsilon}_{\theta}(z_t, \emptyset_{desc}, \mathbf{c}_{cont}) - \boldsymbol{\epsilon}_{\theta}(z_t, \emptyset_{desc}, \emptyset_{cont})\Bigr)
\end{multline}

Derivations for Equation \ref{eq:cfg2} are included in the Appendix\footnote{https://voiceldm.github.io}. When $w_{desc} = w_{cont}$, its effect is equivalent with that of Equation \ref{eq:cfg}.
By appropriately manipulating the values of $w_{desc}$ and $w_{cont}$, one can effectively regulate the guidance strength for each individual condition.
As an example, one may increase the value of $w_{cont}$ but assign a lower value of $w_{desc}$ to obtain audio with increased style diversity while having more linguistic accuracy. An analysis exploring the effect of dual classifier-free guidance is conducted in Section \ref{ssec:cfgeff}. To enable the use of dual classifier-free guidance during inference, we randomly drop the conditions $\mathbf{c}_{desc}$, $\mathbf{c}_{cont}$ independently during training.

\section{Experiment Settings}
\label{sec:expsettings}

\subsection{Data Preparation}
\label{ssec:dataprep}
We use the following publicly available real-world audio datasets for training: 
% (1) AudioSet \cite{gemmeke2017audio}, a large-scale dataset of various general audio extracted from YouTube videos. (2) The English subset of the CommonVoice 13.0 corpus \cite{ardila2020common}, a dataset of real-world speech utterances from various speakers. (3) VoxCeleb1 \cite{nagrani2017voxceleb}, utterances from celebrities extracted from YouTube videos. (4) DEMAND \cite{thiemann2013diverse}, a dataset of various recordings of environmental noises. To prepare the training dataset, we collect English speech segments and non-speech segments from these real-world audio datasets. We include all audios from CommonVoice and VoxCeleb as speech segments and include all audios from DEMAND as non-speech segments.
AudioSet \cite{gemmeke2017audio}, the English subset of the CommonVoice 13.0 corpus \cite{ardila2020common}, VoxCeleb1 \cite{nagrani2017voxceleb}, and DEMAND \cite{thiemann2013diverse}. To prepare the training dataset, we allocate each audio from these real-world audio datasets into either English speech segments or non-speech segments. We include all audios from CommonVoice and VoxCeleb as speech segments and include all audios from DEMAND as non-speech segments.

To process AudioSet, we leverage an automatic speech recognition model Whisper \cite{radford2023robust}, where we use two versions of the model: \textit{large-v2} and \textit{medium.en}. \textit{large-v2} is a multilingual model that also has language identification capabilities, whereas \textit{medium.en} is more specialized in English. 
First we feed all audio into \textit{medium.en} and generate the transcriptions. With the transcriptions from \textit{medium.en}, we classify audio that contains intelligible English speech from those that do not.
To further ensure that the audios are correctly classified, we additionally use \textit{large-v2} for audios that have been classified as speech segments. With the language identification functionality of \textit{large-v2}, for each audio we compute the probability of the language being English and generate the transcriptions. We only classify audio as English speech segments if the probability that the language is English is greater than 50\%, and the word error rate (WER) between the transcriptions of \textit{large-v2} and \textit{medium.en} is less than 50\%.

% Otherwise, they are included as non-speech segments.
% After filtering, there are still inaccurate transcriptions within the speech segments (e.g. transcriptions existing for random noise or non-English speech).
% Thus we conduct additional filtering using \textit{large-v2}.
% For each audio in speech segments, we compute the probability of the language being English and generate speech-to-text transcriptions with \textit{large-v2}.
% We collect audio that has more than 50\% probability of the language being English, and the word-error-rate (WER) between the transcriptions of \textit{large-v2} and \textit{medium.en} is less than 50\%.

After the audios are classified into speech segments and non-speech segments, we use the transcriptions generated by \textit{medium.en} to be used as $text_{cont}$ for every audio in the speech segments. We use the transcriptions from \textit{medium.en} instead of \textit{large-v2} since we find that it generates slightly more accurate transcriptions for general audio such as AudioSet. For audios longer than 10 seconds, we take the first 10 seconds of audio before feeding into \textit{medium.en} to generate the transcriptions. For audios that already have pre-existing transcriptions and has a duration of less than 10 seconds, we use the provided transcriptions to achieve better performance.

In total, 2.43M speech segments and 824k non-speech segments are collected. All audio files are resampled into 16kHz sampling rate and mono format. 
All audios in the speech segments are standardized to have a duration of 10 seconds, either by selecting the initial 10 seconds for longer clips or zero-padding shorter segments.

\subsection{Model Configuration}
We train two models, VoiceLDM-S and VoiceLDM-M. 
The difference between these two models is the size of the U-Net backbone. We use the U-Net used in \cite{liu2023audioldm}, where the channel dimensions of the encoder blocks are $[c_u, 2c_u, 3c_u, 5c_u]$, where $c_u$ is the basic channel number. We use $c_u = 128$ for VoiceLDM-S and $c_u = 192$ for VoiceLDM-M. This results in a total of 280M and 508M number of trainable parameters, including the content encoder and the differentiable durator. To condition the U-Net backbone with two conditions, we replace the self-attention component of the U-Net with cross-attention to additionally condition $\mathbf{c}_{cont}$. $\mathbf{c}_{desc}$ is conditioned in the same way as \cite{liu2023audioldm}, by concatenating it with the timestep embedding.

We employ the pre-trained VAE and vocoder from \cite{liu2023audioldm}. We use the pre-trained CLAP model \cite{wu2023large} released by the authors\footnote{https://huggingface.co/laion/clap-htsat-unfused}. 
% For the content encoder, it is possible to train a Transformer encoder from scratch. However, in pursuit of improved performance, we tear off a Transformer encoder part from a pre-trained SpeechT5 \cite{ao2022speecht5} model trained for speech synthesis\footnote{https://huggingface.co/microsoft/speecht5\_tts}.
For the content encoder, it is possible to train a Transformer encoder from scratch. However, we extract a Transformer encoder component from a pre-trained SpeechT5 \cite{ao2022speecht5} model trained for TTS\footnote{https://huggingface.co/microsoft/speecht5\_tts}, in pursuit of improved performance.

\subsection{Training Configuration}

We use two NVIDIA A5000 GPUs with a batch size of 8 each for VoiceLDM-S and use four NVIDIA A5000 GPUs with a batch size of 4 each to train VoiceLDM-M. Both models are trained for 3M steps.
The learning rate is set to $2e-5$ for the AdamW optimizer.

We use audios from the speech segments to train VoiceLDM. However, if the speech segment is from CommonVoice, randomly selected audio from the non-speech segments is mixed on-the-fly with a probability of 0.5. For non-speech segments, the audio is randomly cut or padded to have a duration of 10 seconds and is mixed with a signal-to-noise ratio (SNR) value randomly selected from a uniform distribution within the range of $[4,~20]$. Otherwise if the speech segment is from AudioSet or VoxCeleb, we do not mix non-speech audio since the audio is already sufficiently noisy.

During training, $\mathbf{c}_{desc}$ and $\mathbf{c}_{cont}$ are randomly dropped with a probability of 0.1 respectively. During inference, we use a DDIM sampler \cite{song2020denoising} with 100 as the number of inference steps.

\subsection{Evaluation Metrics}
We use quantitative and qualitative metrics to assess the audio quality and the input prompt adherence of VoiceLDM.

\noindent\textbf{Quantitative Metrics.} We report Frechet Audio Distance (FAD), Kullback-Leiber (KL) divergence, and CLAP score. Additionally, to evaluate speech intelligibility, we measure the word error rate (WER) with Whisper \textit{large-v2}. We also report the word error rate ($\Delta$WER) between the transcriptions of two Whisper models, \textit{large-v2} and \textit{medium.en}. Having a lower value of $\Delta$WER suggests that the generated audio has high speech intelligibility.

\noindent\textbf{Qualitative Metrics.} We report overall impression (OVL), relevance between audio and condition (REL), and mean opinion score (MOS) of the generated audio. For qualitative evaluation, we use crowd-sourcing and ask participants to rate the audio on a scale between 1 to 5. We make sure each audio is evaluated by at least 10 different raters.

\section{Results}
\label{sec:result}

\subsection{Main Result}

We evaluate the performance of VoiceLDM on the AudioCaps \cite{kim2019audiocaps} test set. Segments containing English speech are collected and the corresponding transcriptions are generated as described in Section \ref{ssec:dataprep}. We denote the original test set as \textit{ac-full} and the processed test set as \textit{ac-filtered}. 
% The number of samples from \textit{ac-filtered} is roughly 22\% of \textit{ac-full}. 
We use the captions from AudioSet as $text_{desc}$ and the generated transcriptions as $text_{cont}$.  We use $w_{desc} = 7, w_{cont} = 7$ for dual classifier-free guidance. We also substitute $text_{desc}$ with the ground truth audio for the descriptive condition, and denote the experiment setting as VoiceLDM-M$_{audio}$. For objective evaluation, we additionally compare VoiceLDM with an AudioLDM 2 \cite{liu2023audioldm2} checkpoint trained for TTS\footnote{We use the \textit{audioldm2-speech-gigaspeech} checkpoint.}, a model also capable of accepting a description and content prompt to generate audio.

\begin{table}[!t]
\vspace{-0.5cm}
\caption{Performance comparison with quantitative metrics on the AudioCaps test set. $\uparrow$: higher is better; $\downarrow$: lower is better.}
\vspace{+0.1cm}
\begin{center}
\resizebox{0.48\textwidth}{!}{%
\begin{tabular}{c|ccc|cc}
\hline
Model & FAD$\downarrow$ & KL$\downarrow$ & CLAP$\uparrow$ & WER(\%)$\downarrow$ & $\Delta$WER(\%)$\downarrow$ \\
\hline
Ground Truth & - & - & 0.251 & 21.21 & 21.21 \\
\hline
VoiceLDM-S & 4.781 & 1.454 & 0.210 & 56.03 & 47.56 \\

VoiceLDM-M & 5.62 & 1.48 & 0.197 & 13.05 & 13.22 \\

VoiceLDM-M$_{audio}$ & 2.499 & 0.883 & 0.209 & 13.41 & 11.82 \\

AudioLDM 2 & 20.720 & 3.005 & 0.060 & 32.84 & 27.39 \\
% Check again
\hline
\end{tabular}
}
\end{center}
\vspace{-0.5cm}
\label{table:mainobj}
\end{table}

\begin{table}[!t]
\vspace{-0.5cm}
\caption{Performance comparison with qualitative metrics on the AudioCaps test set. We report overall quality (OVL), relevance between the audio and descriptive prompt (REL$_{desc}$), and the relevance between the audio and content prompt (REL$_{cont}$).}
\vspace{+0.1cm}
\begin{center}
\resizebox{0.4\textwidth}{!}{%
\begin{tabular}{c|ccc}
\hline
Model & OVL & REL$_{desc}$ & REL$_{cont}$ \\
\hline
Ground Truth & 4.27 & 4.30 & 4.45 \\
\hline
VoiceLDM-S & 3.61 & 3.66 & 4.14 \\

VoiceLDM-M & 3.88 & 3.89 & 4.52 \\

VoiceLDM-M$_{audio}$ & 4.08 & 4.03 & 4.61 \\
\hline
\end{tabular}
}
\end{center}
\vspace{-0.5cm}
\label{table:mainsub}
\end{table}

Quantitative and qualitative evaluation results are shown in Table \ref{table:mainobj} and Table \ref{table:mainsub}. VoiceLDM is capable of generating audio that adheres to both input conditions simultaneously. The largest model VoiceLDM-M, even surpasses the speech intelligibility of the ground truth audio, while maintaining competitive audio quality and description prompt adherence. Substituting $text_{desc}$ with the audio yields improved outcomes in following the environmental context, while also achieving high speech intelligibility. 
AudioLDM 2, a TTS-focused model, fails to adhere to the given description prompt if the prompt encompasses more than just speech-related elements.

\subsection{Text-to-Speech Capabilities}

VoiceLDM has the ability to act as a regular TTS model with the prompt \textit{``clean speech"} as input for $text_{desc}$. We evaluate the TTS capabilities of VoiceLDM on the CommonVoice test set. The transcriptions from the CommonVoice test set are given as $text_{cont}$. We use $w_{desc} = 1, w_{cont} = 9$ for dual classifier guidance. We compare the performance with SpeechT5 trained for TTS and FastSpeech 2 trained on CommonVoice\footnote{https://huggingface.co/facebook/fastspeech2-en-200\_speaker-cv4}.

\begin{table}[!t]
% \vspace{-0.5cm}
\caption{Performance comparison on TTS capabilities on the CommonVoice test set.}
\vspace{+0.1cm}
\begin{center}
\resizebox{0.425\textwidth}{!}{%
\begin{tabular}{c|cc|c}
\hline
Model & WER(\%)$\downarrow$ & $\Delta$WER(\%)$\downarrow$ & MOS \\
\hline
Ground Truth & 11.818 & 11.818 & 4.09 \\
\hline
VoiceLDM-S & 10.693 & 8.378 & 3.74 \\

VoiceLDM-M & 3.909 & 2.390 & 3.96 \\

VoiceLDM-M$_{audio}$ & 10.459 & 7.911 & 3.89 \\

FastSpeech 2 \cite{ren2020fastspeech} & 9.751 & 9.327 & 3.32 \\

SpeechT5 \cite{ao2022speecht5} & 6.384 & 3.054 & 3.74 \\
\hline
\end{tabular}
}
\end{center}
\vspace{-0.5cm}
\label{table:tts}
\end{table}

Table \ref{table:tts} shows the results of TTS evaluation. Evaluation on the CommonVoice test set reveals that all VoiceLDM models are able to surpass the ground truth audio in terms of linguistic intelligibility, as measured by WER and $\Delta$WER.
The largest model, VoiceLDM-M achieves the lowest WER and $\Delta$WER and even achieves naturalness comparable to ground truth audio. VoiceLDM-M also outperforms FastSpeech 2 and SpeechT5 across all metrics by a significant margin.

\subsection{Text-to-Audio Capabilities}
Although VoiceLDM is trained solely on audio samples with human voices, it exhibits the ability to perform regular zero-shot TTA.
We evaluate the zero-shot TTA capabilities of VoiceLDM on the AudioCap test set. The captions provided from the AudioCap test set are given as $text_{desc}$, and an empty string is given as $text_{cont}$. $w_{desc} = 9, w_{cont} = 1$ is used for dual classifier guidance.

\begin{table}[!t]
\vspace{-0.5cm}
\caption{Performance comparison on TTA capabilities on the AudioCaps test set.}
% \vspace{+0.1cm}
\begin{center}
\resizebox{0.335\textwidth}{!}{%
\begin{tabular}{c|ccc}
\hline
Model & FAD$\downarrow$ & KL$\downarrow$ & CLAP$\uparrow$  \\
\hline
\textit{ac-full} &  &  &  \\
Ground Truth & - & - & 0.259 \\
VoiceLDM-S & 15.073 & 3.309 & 0.089 \\
VoiceLDM-M & 10.119 & 2.458 & 0.172 \\
AudioLDM-S & 5.131 & 1.823 & 0.178 \\
AudioLDM-M & 4.689 & 1.986 & 0.224 \\
\hline
\textit{ac-filtered} &  &  &  \\
Ground Truth & - & - & 0.247 \\
VoiceLDM-S & 9.852 & 2.504 & 0.099 \\
VoiceLDM-M & 6.819 & 1.713 & 0.185 \\
AudioLDM-S & 3.211 & 1.485 & 0.175 \\
AudioLDM-M & 2.974 & 1.621 & 0.215 \\
\hline
\end{tabular}
}
\end{center}
\vspace{-0.5cm}
\label{table:tta}
\end{table}

The results in Table \ref{table:tta} show that despite not being specifically trained for TTA, VoiceLDM is capable of generating plausible audio as seen in TTA models. VoiceLDM-M achieves comparable results in terms of KL and CLAP scores when compared with AudioLDM-S, a model specifically trained for TTA. The gap in performance becomes smaller when evaluated on the \textit{ac-filtered} test set, even outperforming AudioLDM-S in terms of CLAP score.

\subsection{Effect of Dual Classifier-Free Guidance}
\label{ssec:cfgeff}

We conduct a series of experiments on the \textit{ac-filtered} test set to explore the effect of dual classifier-free guidance. We compare the performance of VoiceLDM-M by only adjusting the values of $w_{desc}$ and $w_{cont}$.

\begin{table}[!t]
% \vspace{-0.5cm}
\caption{Effect of dual classifier-free guidance.}
\vspace{+0.3cm}
\begin{center}
\resizebox{0.425\textwidth}{!}{%
\begin{tabular}{cc|ccc|c}
\hline
$w_{desc}$ & $w_{cont}$ & FAD$\downarrow$ & KL$\downarrow$ & CLAP$\uparrow$ & WER(\%) \\
\hline
5 & 5 & 5.08 & 1.45 & 0.193 & 15.95 \\

5 & 7 & 5.60 & 1.53 & 0.172 & 13.95 \\

5 & 9 & 6.15 & 1.69 & 0.170 & 12.41 \\

7 & 5 & 5.25 & 1.42 & 0.198 & 17.67 \\

7 & 7 & 5.62 & 1.48 & 0.197 & 13.05 \\

7 & 9 & 6.16 & 1.54 & 0.175 & 12.65 \\

9 & 5 & 5.50 & 1.44 & 0.196 & 18.46 \\

9 & 7 & 5.81 & 1.47 & 0.190 & 13.63 \\

9 & 9 & 6.14 & 1.54 & 0.177 & 13.97 \\
\hline
\end{tabular}
}
\end{center}
\vspace{-0.5cm}
\label{table:dualcfg}
\end{table}

As shown in Table \ref{table:dualcfg}, while using a high value of $w_{cont}$ yields high speech intelligibility, it leads to a trade-off in reduced adherence to the description prompt. 
% The opposite is also true for $w_{desc}$ as well.
Conversely, increasing $w_{desc}$ enhances adherence but compromises speech intelligibility.
Adjusting the value of $w_{desc}$ and $w_{cont}$ allows one to balance this trade-off, thereby facilitating the generation of more desirable outcomes.

\section{Conclusion}
\label{sec:foot}

This paper introduces VoiceLDM, a model that introduces unique functionality to control TTS generation with environmental context. VoiceLDM is trained with vast quantities of real-audio data through the utilization of CLAP and Whisper. We improve model controllability by employing dual classifier-free guidance, which enables one to control the trade-off of the guidance strength for each condition. Quantitative and qualitative evaluation results show that VoiceLDM is simultaneously capable of achieving the speech synthesis and general audio synthesis functionalities found in TTS and TTA models. Furthermore, we show that VoiceLDM can function as a conventional TTS or TTA model, positioning itself as a generalized extension of the two domains.

%\vfill
%\pagebreak

\vfill\pagebreak

\bibliographystyle{IEEEtran}
% \bibliographystyle{IEEEbib}
% \bibliography{shortstrings,main}

\vfill\pagebreak

\appendix
\section{Derivation of Dual Classifier-Free Guidance}
\label{app:cfgproof}

% $\boldsymbol{\epsilon} \propto 

% Given two conditions $c_1$, $c_2$, let $\nabla_{z_t}\log p_{\theta}(z_t|c_1, c_2)$ be the gradient of the log-density of the conditional distribution of $z_t$, predicted by the noise prediction network $\theta$.

Given two conditions $c_1$, $c_2$, let $p_{\theta}(z_t|c_1, c_2)$ be the density of the conditional distribution of $z_t$, which is estimated by a score prediction network $\theta$.
When applying classifier-free guidance \cite{ho2021classifier} for two conditions, the conditional distribution of $p_{\theta}$ is modified with additional guidance with strength $w$ as follows:

\begin{equation}
    \Tilde{p}_{\theta}(z_t | c_1, c_2) \propto p_{\theta}(z_t | c_1, c_2)p_{\theta}(c_1, c_2 | z_t)^w
\end{equation}

For the case of VoiceLDM, it is reasonable to assume that the two conditions $c_1$ and $c_2$ are independent. In this case, the conditional distribution is modified as follows:

\begin{equation}
    \Tilde{p}_{\theta}(z_t | c_1, c_2) \propto p_{\theta}(z_t | c_1, c_2)p_{\theta}(c_1 | z_t)^w p_{\theta}(c_2 | z_t)^w
\end{equation}

One may also consider the possibility of using different guidance strengths for each condition, where we denote the individual guidance strengths as $w_1$ and $w_2$:

\begin{equation}
    \Tilde{p}_{\theta}(z_t | c_1, c_2) \propto p_{\theta}(z_t | c_1, c_2)p_{\theta}(c_1 | z_t)^{w_1} p_{\theta}(c_2 | z_t)^{w_2}
\end{equation}

From this we get the gradient of the log-density of the modified conditional distribution as

\begin{align}
    & \nabla_{z_t}\log \Tilde{p}_{\theta}(z_t|c_1, c_2) \nonumber \\ 
    & = \nabla_{z_t}\log p_{\theta}(z_t|c_1, c_2)p_{\theta}(c_1 | z_t)^{w_1} p_{\theta}(c_2 | z_t)^{w_2} \nonumber \\
    & = \nabla_{z_t}\log p_{\theta}(z_t|c_1, c_2) \biggl(\frac{p_{\theta}(z_t|c_1)}{p_{\theta}(z_t)}\biggr)^{w_1} \biggl(\frac{p_{\theta}(z_t|c_2)}{p_{\theta}(z_t)}\biggr)^{w_2} \nonumber \\
    & = \nabla_{z_t}\log p_{\theta}(z_t|c_1, c_2) \nonumber \\ 
    & \quad + w_1\Bigl(\nabla_{z_t}\log p_{\theta}(z_t|c_1) - \nabla_{z_t}\log p_{\theta}(z_t)\Bigr) \nonumber \\ 
    & \quad + w_2\Bigl(\nabla_{z_t}\log p_{\theta}(z_t|c_2) - \nabla_{z_t}\log p_{\theta}(z_t)\Bigr)
\end{align}

Finally, this can be rewritten in terms of diffusion scores:

\begin{align}
    \boldsymbol{\Tilde{\epsilon}}_{\theta}(z_t, c_1, c_2) &= \boldsymbol{\epsilon}_{\theta}(z_t, c_1, c_2) \nonumber \\
    & \quad + w_1\Bigl(\boldsymbol{\epsilon}_{\theta}(z_t, c_1) - \boldsymbol{\epsilon}_{\theta}(z_t) \Bigr) \nonumber \\
    & \quad + w_2\Bigl(\boldsymbol{\epsilon}_{\theta}(z_t, c_2) - \boldsymbol{\epsilon}_{\theta}(z_t) \Bigr)
\end{align}

\end{document}